\documentclass[aps,prd,10pt,twocolumn,showpacs,superscriptaddress]{revtex4-1}

\usepackage{amsmath,amssymb,amsfonts}
\usepackage{slashed}

\usepackage{graphicx}
\usepackage{color}
\usepackage{hyperref}
\definecolor{darkred}{rgb}{0.8,0.1,0.1}
\hypersetup{
    colorlinks=true,
    linkcolor=darkred, 
    citecolor=blue, 
}

\newcommand{\R}{\ensuremath{\mathcal R}}
\newcommand{\D}{\ensuremath{\mathcal D}}
\newcommand{\bpsi}{\ensuremath{\overline\psi{}}}
\newcommand{\bPsi}{\ensuremath{\overline\Psi{}}}
\newcommand{\Sol}{\mathsf{Sol}}

\begin{document}

\title{Quantization of the massive gravitino on FRW spacetimes}

\author{Alexander Schenkel}
\email{schenkel@math.uni-wuppertal.de}
\affiliation{Fachgruppe Mathematik\\
Bergische~Universit\"at~Wuppertal,~Gau\ss stra\ss e~20,~42119~Wuppertal,~Germany\\[2mm]}
\affiliation{Institut f\"ur Theoretische Physik und Astrophysik\\
  Universit\"at W\"urzburg, Am Hubland, 97074 W\"urzburg, Germany\\[-2mm] \hfill}

\author{Christoph F.~Uhlemann}
\email{uhlemann@physik.uni-wuerzburg.de}

\affiliation{Institut f\"ur Theoretische Physik und Astrophysik\\
  Universit\"at W\"urzburg, Am Hubland, 97074 W\"urzburg, Germany\\[-2mm] \hfill}

\date{\today}



\begin{abstract}
In this article we study the quantization and causal properties of a massive
spin $3/2$ Rarita-Schwinger field on spatially flat Friedmann-Robertson-Walker (FRW) spacetimes.
We construct Zuckerman's universal conserved current
and prove that   
it leads to a positive definite inner product on solutions of the field equation.
Based on this inner product, we quantize the Rarita-Schwinger field in terms of a CAR-algebra.
The transversal and longitudinal parts constituting the independent on-shell degrees of freedom decouple.
We find a Dirac-type equation for the transversal polarizations, ensuring a causal propagation.
The equation of motion for the longitudinal part is also of Dirac-type, but with respect to
an `effective metric'.
We obtain that 
for all four-dimensional FRW solutions with a matter equation of state $p =\omega \rho$ and $\omega\in(-1,1]$
the light cones of the effective metric are more narrow than the 
standard cones, which are recovered for the de Sitter case $\omega=-1$.
In particular, this shows that the propagation of the longitudinal part, 
although non-standard for $\omega\neq -1$, is completely causal in cosmological constant, 
dust and radiation dominated universes.
\end{abstract}


\maketitle
 

\section{Introduction}

Supergravity is a well-motivated extension of Einstein's theory of general relativity 
and may have interesting consequences for cosmology and particle physics.
Specific issues like production mechanisms and properties of gravitino dark matter or
the analysis of scattering experiments are best 
addressed in terms of effective quantum field theory for the fluctuations around 
appropriate solutions of classical supergravity.
While this means Minkowski-space for collider physics, 
the less symmetric FRW backgrounds are of particular interest for studies of the early universe.
A consistent quantum field theory for the gravitino linearized around such supergravity solutions
is therefore of great importance for physical applications.

A matter-coupled supergravity with FRW solutions has been proposed and analyzed in detail 
at the classical level by Kallosh, Kofman, Linde and Van Proeyen \cite{Kallosh:1999jj,Kofman:1999ip,Kallosh:2000ve}.
It has been shown that due to the special form of the scalar field potential, 
the massive gravitino equations are consistent and the propagation is causal on the FRW solutions.
However, the quantization of the massive gravitino on spacetimes which are not 
necessarily Einstein  has been discussed only recently \cite{Hack:2011yv},
arriving at the conclusion that a consistent quantization is
only possible on Einstein spaces, i.e.~when the Einstein tensor is proportional to the metric. 
This statement is based on the non-conservation of the specific gravitino current proposed in \cite{Hack:2011yv},
which would lead to inconsistencies when imposing canonical anticommutation relations (CAR).

In this work we improve on this point and show that a consistent quantization of the gravitino 
on FRW spacetimes is indeed possible.
We consider the Rarita-Schwinger field in $d$ dimensions without assuming a specific model,
but
with properties general enough to include the relevant supergravity cases.
In particular, we allow for a spacetime-dependent mass as it arises in linearizations of supergravity
around non-trivial backgrounds like  FRW \cite{Kallosh:1999jj,Kofman:1999ip,Kallosh:2000ve}, 
but do not fix the dependence a priori. 
We show that there is a canonical conserved current for the Rarita-Schwinger field
on all spacetimes of dimension $d\geq 3$.
Specializing to the case of $d$-dimensional spatially flat
FRW spacetimes, we prove that the inner product derived from this current is positive definite
on solutions of the Rarita-Schwinger equation.
In particular, it satisfies non-negativity, the necessary condition for a consistent implementation
of CAR emphasized in \cite{Hack:2011yv}.
We construct the CAR-algebra and discuss causality and the role of supergravity in that respect.
We find that the propagation is in general non-standard, yet completely causal on a wide class of FRW spacetimes
including dust, radiation and cosmological constant dominated universes.
Specifically, the domains of dependence on these spacetimes are 
in general more narrow than na\"{\i}vely expected.
Time-variations of the mass stretch these domains, eventually arriving at the standard light cones
 in the supergravity model of \cite{Kallosh:1999jj,Kofman:1999ip,Kallosh:2000ve}.

The outline is as follows: In Section \ref{sec:eom} we review the action, equation of motion and constraints
for a massive Rarita-Schwinger field on a $d$-dimensional spacetime.
We derive a current for this field in Section \ref{sec:current}
using variational bicomplex methods \cite{0669.58014,0966.81533,Reyes:2004zz}
and show that this current is conserved.
For $d$-dimensional spatially flat FRW spacetimes we prove in Section \ref{sec:innerproduct} 
that the derived inner product is positive when evaluated on solutions of the equation of motion.
The quantization is outlined in Section \ref{sec:quant} and a discussion of the causal propagation
and the role of supersymmetry is given in Section
\ref{sec:causal}. 
Examples of cosmological spacetimes allowing for a causal propagation are studied in Section \ref{sec:examples}.
We conclude in Section \ref{sec:conc}.
Our notation and conventions are summarized in 
Appendix~\ref{app:notation}.


\section{\label{sec:eom}Equation of motion and constraints}

We consider a Dirac Rarita-Schwinger field $\psi^{}_\mu$ on a spacetime of dimension $d\geq 3$
with metric signature mostly minus.
The action reads
\begin{flalign}\label{eqn:gravitino-action}
 S = \int d^dx\, e ~\bpsi^{}_\mu \R^\mu[\psi]~,
\end{flalign}
with the Rarita-Schwinger operator
\begin{flalign}\label{eqn:rsoperator}
\R^\mu[\psi]:=i\gamma^{\mu\nu\rho}\D_\nu\psi_\rho+m\gamma^{\mu\nu}\psi_\nu~.
\end{flalign}
The covariant derivative is 
$\D_\mu\psi_\nu :=\partial_\mu\psi_\nu+\frac{1}{4}\omega_{\mu ab}\gamma^{ab}\psi_\nu-\Gamma_{\mu\nu}^\rho\psi_\rho$.
We assume a torsion-free background configuration, i.e.~the connection symbols are symmetric
${\Gamma_{\mu\nu}^\rho = \Gamma_{\nu\mu}^\rho}$.
The mass $m$ may be spacetime-dependent, but is assumed to be real and positive.
This action with $d{=}4$ is the quadratic gravitino part of the matter-coupled $\mathcal{N}{=}1$
supergravity discussed in \cite{Kallosh:1999jj,Kofman:1999ip,Kallosh:2000ve},
up to metric conventions and the Majorana condition.
The mass $m$ in \cite{Kallosh:1999jj,Kofman:1999ip,Kallosh:2000ve} is related to the K\"ahler and superpotential via
$m=e^{K/2}\,W/M_\mathrm{P}^2$.
The action (\ref{eqn:gravitino-action}) is real up to a boundary term and 
the Rarita-Schwinger operator (\ref{eqn:rsoperator}) is formally
self-adjoint with respect to $\bigl(\psi^{}_1,\psi^{}_2\bigr):=\int d^dx\, e~\bpsi_1^\mu\psi^{}_{2\mu}$.
That is, for all $\psi^{}_1$ and $\psi^{}_2$ with supports of compact overlap we have
\begin{flalign}
 \bigl(\psi^{}_1,\R[\psi^{}_2]\bigr) = \bigl(\R[\psi^{}_1],\psi^{}_2\bigr)~.
\end{flalign}
Contracting the equation of motion $\R^\mu[\psi]=0$ with $\gamma_\mu$ 
leads to the on-shell constraint
\begin{subequations}\label{eqn:constr}
\begin{flalign}\label{eqn:constr1}
 i\, \slashed{\D}\, \gamma\cdot\psi - i\,\D_\mu\psi^\mu + \frac{d-1}{d-2}\,m~\gamma\cdot\psi=0~.
\end{flalign}
Acting with $\D_\mu$ on $\R^\mu[\psi]=0$ and using (\ref{eqn:constr1}) yields the second constraint
\begin{flalign}\label{eqn:constr2}
 \frac{i}{2} G^{\mu\nu}\gamma^{}_\mu\psi^{}_\nu +(\partial_\mu m) \,\gamma^{\mu\nu}\psi^{}_\nu + \frac{d-1}{d-2} i\,m^2\,\gamma\cdot\psi=0~,
\end{flalign}
\end{subequations}
where $G^{\mu\nu} := \R^{\mu\nu}-\frac{1}{2}g^{\mu\nu}\,\R$ is the Einstein tensor.
Using (\ref{eqn:constr1}), the Rarita-Schwinger equation $\R^\mu[\psi]=0$ can be written as
\begin{flalign}\label{eqn:gravitino-eom2}
\big(i\,\slashed{\D}-m\big)\psi^{}_\mu - \big(i\,\D_\mu+\frac{m}{d-2}\,\gamma^{}_\mu\big)\gamma\cdot\psi=0~.
\end{flalign}
Due to the derivative in the second term, (\ref{eqn:gravitino-eom2}) is not of Dirac-type \cite{Bar:2011iu}
and the causal propagation of the Rarita-Schwinger field on a generic spacetime is not guaranteed a priori.
We will discuss this point further in Section \ref{sec:causal} and \ref{sec:examples}.


\section{\label{sec:current}Conserved current}

We construct Zuckerman's universal conserved current \cite{0669.58014,0966.81533,Reyes:2004zz}
for the Rarita-Schwinger field using the variational bicomplex.
We will verify its conservation explicitly, so the reader may also pass directly to (\ref{eqn:current-one-form}).

The basic idea of the variational bicomplex is to consider functions and differential forms
on the product space $\mathcal{M}\times \mathcal{S}$, with $\mathcal{M}$ being spacetime and $\mathcal{S}$
the space of field configurations (this can be made precise by using $\infty$-jet bundles \cite{Reyes:2004zz}).
The differential forms on $\mathcal{M}\times \mathcal{S}$ can be decomposed into subspaces of a definite
horizontal (i.e.~spacetime) and vertical (i.e.~field space) degree. 
Likewise, the exterior differential on $\mathcal{M}\times \mathcal{S}$ splits into
a horizontal differential $\mathrm{d}$ and a vertical differential $\delta$, increasing the horizontal/vertical
degree by one.

The starting point of the construction is a Lagrangian described by a $(d,0)$-form, i.e.~of maximal horizontal degree,
on $\mathcal{M}\times \mathcal{S}$. 
The Lagrangian form corresponding to the Dirac Rarita-Schwinger action (\ref{eqn:gravitino-action}) reads
\begin{flalign}\label{eqn:lagform}
 L=i\,\bpsi\wedge\star V^3\wedge \D \psi + (-1)^d m\,\bpsi\wedge\star V^2\wedge\psi
~,
\end{flalign}
where $\star$ denotes the Hodge operator, $\psi:=\psi^{}_\mu dx^\mu$ and
$\D\psi := \D_\mu \psi^{}_\nu \,dx^\mu\wedge dx^\nu$. 
Furthermore, ${V:=\gamma_\mu dx^\mu}$ and the normalized $n$-fold product
is denoted by ${V^n:=\frac{1}{n!} V\wedge..\wedge V}$.
The vertical exterior derivative of the Lagrangian admits a decomposition
\begin{flalign}
 \delta L=E+\mathrm{d}\Theta~,
\end{flalign}
with a unique source form $E$ of degree $(d,1)$
yielding the equations of motion
and $\Theta$ of degree $(d{-}1,1)$, which is unique up to horizontally exact parts.
For the Lagrangian (\ref{eqn:lagform}) we find
\begin{flalign}\label{eqn:theta}
 \Theta = -i\, \bpsi\wedge \star V^3\wedge\delta\psi~.
\end{flalign}
Zuckerman's universal current is defined as the contraction of the 
$(d{-}1,2)$-form $\mathsf{u}:=\delta\Theta$ with two Jacobi fields, i.e.~solutions of the linearized equations of motion.
Since we are considering a linear theory, the Jacobi fields coincide with solutions of the Rarita-Schwinger equation
$\R_{}^\mu[\psi]=0$.
From (\ref{eqn:theta}) we find
\begin{flalign}
\mathsf{u} = -i\delta\bpsi\wedge\star V^3\wedge \delta\psi~,
\end{flalign}
and contracting with the two Jacobi fields
$\bpsi_1$ and $\psi^{}_2$ we obtain the $(d{-}1,0)$-form current
\begin{flalign}\label{ref:Ucontr}
 \mathsf{u}[\bpsi^{}_1,\psi^{}_2]=i(-1)^d\,\bpsi^{}_1\wedge\star V^3\wedge \psi^{}_2~.
\end{flalign}
Note that (\ref{ref:Ucontr}) does not depend on the field space coordinates. 
We pull back (\ref{ref:Ucontr}) to $\mathcal{M}$ to obtain a $d{-}1$-form current (denoted by the same symbol) on spacetime.
From that current on $\mathcal{M}$ we
define the more familiar one-form current 
 ${j[\bpsi_1,\psi_2]:=i\star\mathsf{u}[\bpsi_1,\psi_2]}$, which
  reads explicitly
\begin{flalign}\label{eqn:current-one-form}
 j_\mu[\bpsi_1,\psi_2] = -\bpsi^{\nu}_{1}\gamma_{\nu\mu\rho}\psi_{2}^{\rho}~.
\end{flalign}
Conservation of the $d{-}1$-form current $\mathsf{u}[\bpsi_1,\psi_2]$, i.e.\ $\mathrm{d}\mathsf{u}[\bpsi_1,\psi_2]=0$, 
is equivalent to $\nabla_\mu j^\mu[\bpsi_1,\psi_2]=0$, with $\nabla_\mu$ being the covariant derivative on vector fields.
We obtain
\begin{flalign}
\nonumber -\nabla_\mu j^\mu[\bpsi_1,\psi_2] 
&= \overline{\D_\mu\psi^{}_{1\nu}}\gamma^{\nu\mu\rho}\psi^{}_{2\rho} 
  +\bpsi^{}_{1\nu} \D_\mu \bigl(\gamma^{\nu\mu\rho}\psi^{}_{2\rho} \bigr)
\\
\nonumber &=\overline{\gamma^{\rho\mu\nu}\D_\mu\psi^{}_{1\nu}} \psi^{}_{2\rho}  + \bpsi^{}_{1\nu}  \gamma^{\nu\mu\rho}\D_\mu\psi^{}_{2\rho}
\\
&=i\,\overline{\R_{}^\rho[\psi_1]} \psi^{}_{2\rho}  -i\,\bpsi^{}_{1\nu}\R_{}^\nu[\psi_2]~.
\end{flalign}
In the first line we have used the Leibniz rule for the covariant derivative 
and $\D_\mu\bpsi_\nu = \overline{\D_\mu\psi_\nu}$,
and in line two that due to the vielbein postulate $\D_\mu\gamma^{\nu\mu\rho}=0$.
Thus, the current is conserved when evaluated on solutions.


\section{\label{sec:innerproduct}Positivity of the inner product}

As noted in \cite{Hack:2011yv}, non-negativity of the inner product constructed from 
the current (\ref{eqn:current-one-form}) is a necessary condition for 
a consistent implementation of CAR.
This is due to the anticommutator of the smeared quantum fields $\Psi_\mu(\bar f^\mu)$ and $\bPsi^\mu(f_\mu)$
being an expression of the form $A^\dagger A+A A^\dagger$, which has a non-negative
expectation value in any normalized state, see also Section \ref{sec:quant} for more details.

We define the inner product associated to (\ref{eqn:current-one-form}) by
\begin{flalign}\label{eqn:innerproduct}
 \langle \psi_1,\psi_2\rangle:=\int_\Sigma n^\mu j_\mu[\bpsi_1,\psi_2]~,
\end{flalign}
where $\Sigma$ is a Cauchy surface with future-directed unit normal vector field $n^\mu$.
Splitting $\mu=(0,m)$ we choose coordinates such that $ds^2=g^{}_{00} d\tau^2+g^{}_{mn}dx^{m}dx^{n}$
and likewise fix $e_0^a=\sqrt{g^{}_{00}}\delta_0^a$.
With a choice of $\Sigma$ such that $n=\sqrt{g^{00}}\partial_\tau$
the integrand evaluates to
\begin{flalign}\label{eqn:integrand}
  n^\mu j_\mu[\bpsi_1,\psi_2]=-\Big(
      \psi_{1m}^\dagger \psi_{2}^{\vphantom{\dagger}m} 
+ \big(\gamma^{m}\psi^{}_{1m}\big)^\dagger\,\big(\gamma^{n}\psi^{}_{2n}\big)\Big)~.
\end{flalign}

We verify non-negativity of the inner product (\ref{eqn:innerproduct})
evaluated on solutions of the Rarita-Schwinger equation
for $d$-dimensional FRW spacetimes, $e_{\mu}^{a}=a(\tau)\delta_{\mu}^a$. 
For compatibility with the FRW symmetries we assume that the mass depends on time only, $m=m(\tau)$.
The spin connection is given by $\omega_{\mu ab}=2  a^\prime a^{-1} e_{\mu[a} e^{0}_{b]}$, 
where prime denotes the derivative with respect to $\tau$.
The constraints (\ref{eqn:constr}) read for the FRW background
\begin{subequations}\label{eqn:constraintsFRW}
\begin{flalign}
\begin{split}\label{eqn:constraint1FRW}
 i\gamma^{\mu\nu}\partial_\mu\psi^{}_\nu 
 + \Big[\frac{i}{2}(d-2)\frac{a^\prime}{a}\gamma^0 +\frac{d-1}{d-2}m \Big] \gamma\cdot\psi
 &\\ -i \Big(d-\frac{3}{2}\Big)\frac{a^\prime}{a}\psi^0&=0~,
\end{split}
\end{flalign}
\begin{flalign} \label{eqn:constraint2FRW}
 \gamma^0\psi^{}_0 = \underbrace{\frac{p-2m^2 \frac{d-1}{d-2} + 
 2i m^\prime\,\gamma^0}{\rho +2 m^2\frac{d-1}{d-2}}}_{=:\,\mathcal{A}}\,\gamma^{m}\psi^{}_{m} ~,
\end{flalign}
\end{subequations}
where for the second equation we have used Friedmann's equations 
$G_0^0 = \rho$ and $G_{m}^{n} =-p\,\delta_{m}^{n}$
(in units $M_{\mathrm{P}}=1$).
These expressions in $d{=}4$ have been obtained in \cite{Kallosh:1999jj,Kofman:1999ip,Kallosh:2000ve}, up to metric conventions.
Combining the $\mu=0$ component of (\ref{eqn:gravitino-eom2}) with (\ref{eqn:constraint1FRW}) yields
\begin{flalign}\label{eqn:constraintderived}
 i\gamma^{mn}\partial^{}_{m}\psi^{}_{n}=-
\Big(\underbrace{m+\frac{i}{2}(d-2)\frac{a^\prime}{a}\gamma^0}_{=:\mathcal B}\Big)\gamma^{m}\psi^{}_{m}~.
\end{flalign}
Due to the constraints \eqref{eqn:constraintderived}, \eqref{eqn:constraint2FRW}, 
only $(d-2)\cdot2^{\lfloor d/2\rfloor}$ of the $d\cdot 2^{\lfloor d/2\rfloor}$ complex degrees of freedom 
of the Rarita-Schwinger field are independent.
It is convenient to transform to spatial Fourier space via 
$\psi^{}_m(\tau,x) = (2\pi)^{1-d}\int d^{d-1}k \:e^{ik_n x^n}\, \widetilde\psi^{}_m(\tau,k)$.
As in \cite{Corley:1998qg,Kallosh:1999jj,Kofman:1999ip,Kallosh:2000ve} 
we separate the spatial part of the Rarita-Schwinger field $\widetilde\psi^{}_m$ into the 
$\gamma_{m}$ and $k_{m}$ traceless part $\widetilde\psi_{m}^\text{\,T}$ and the traces  
\begin{flalign}
 \widetilde\chi:=\gamma_{}^n\widetilde\psi^{}_n~,
 \qquad \widetilde\zeta:=k^n\widetilde\psi^{}_n~.
\end{flalign}
With $\hat k_{m}:=k_{m}/\vert k\vert$, $\vert k\vert := \sqrt{-k_{n}k^{n}}$ and 
$\hat{\slashed{k}}:=\hat k_{m}\gamma^{m}$ we find
\begin{flalign}\label{eqn:decomposition}
 \widetilde\psi^{}_{m}=\widetilde\psi_{m}^\text{\,T}
 +\frac{\gamma^{}_{m}+\hat k^{}_{m}\,\hat{\slashed{k}} }{d-2}\:\widetilde\chi
 +\frac{\gamma^{}_{m}\hat{\slashed{k}} -(d-1)\hat k^{}_{m}}{(d-2)|k|}\:\widetilde\zeta~.
\end{flalign}
The constraint (\ref{eqn:constraintderived}) in $k$-space yields the following relation between 
the traces
\begin{flalign}\label{eqn:constraintkspace}
 \widetilde\zeta = 
\big(\slashed{k}-\mathcal B\big)\,\widetilde\chi~.
\end{flalign}
With (\ref{eqn:constraintkspace}) the decomposition (\ref{eqn:decomposition}) becomes
\begin{flalign}\label{eqn:decomposition2}
 \widetilde\psi_{m} = \widetilde\psi_{m}^\text{\,T} -\bigg(\hat k_{m} \hat{\slashed{k}} -
\frac{(d-1)\hat k_{m} -\gamma_{m}\hat{\slashed{k}}}{(d-2)\vert k\vert} \,\mathcal B \bigg)\,\widetilde\chi~.
\end{flalign}
The traceless part $\varPsi_{m}^\text{T}$ comprises 
$(d-3)\cdot2^{\lfloor d/2\rfloor}$ degrees of freedom and the trace part $\gamma^{n}\varPsi^{}_{n}$
the remaining $2^{\lfloor d/2\rfloor}$.
Using this on-shell decomposition, the inner product in Fourier space evaluates to 
\begin{flalign}\label{eqn:innerproductx}
 \langle \psi_1,\psi_2\rangle = \int\frac{d_{}^{d-1}k}{(2\pi)^{d-1}}\, a^{d-1} 
\left(-\widetilde\psi_{1m}^{\text{\,T}\,\dagger}\widetilde\psi_{2}^{\vphantom{\dagger}\text{\,T}m} + \mathcal{C}\, \widetilde\chi_1^{\,\dagger}\widetilde\chi^{\vphantom{\dagger}}_2\right)~,
\end{flalign}
where 
\begin{flalign}
 \mathcal{C} = \frac{d-1}{(d-2)~ \vert k \vert^2}  \Big(m^2 +\frac{1}{4}(d-2)_{}^2\, \frac{{a^\prime}^2}{a^4}\Big)
\end{flalign}
is positive.
The integrand is pointwise (in $k$-space) non-negative, since in our conventions the spatial metric is negative definite.
Thus, for any nonzero solution $\psi\not\equiv 0$ the norm is positive,  $\langle \psi,\psi\rangle >0$.

\section{\label{sec:quant}Quantization}

Using the inner product (\ref{eqn:innerproduct}), we can quantize the Dirac Rarita-Schwinger
field on $d$-dimensional FRW spacetimes analogously to the spin $1/2$ Dirac field \cite{0518.58018}.
We briefly outline the construction of the CAR-algebra and refer for
details on fermionic quantization to \cite{0518.58018,Dappiaggi:2009xj,Bar:2011iu} and references therein.

We denote by $\Sol$ the space of spinor solutions of the Rarita-Schwinger equation
 which are of compact support when restricted to any Cauchy surface.
The space $\overline{\Sol}$ of cospinor solutions is defined as the image of $\Sol$ under the map
$\Sol\ni f_\mu\mapsto \overline{f}{}^\mu$. 
To the spinor/cospinor solutions we associate smeared field operators
via $\mathbb{C}$-linear maps $f_\mu\mapsto \bPsi^\mu(f_\mu)$ and
 $\overline{f}{}^\mu\mapsto \Psi_\mu(\overline{f}{}^\mu)$. 
The CAR-algebra is defined as the $\ast$-algebra with unit $\mathsf{1}$ generated by these operators,
subject to the relations
\begin{subequations}\label{eqn:CAR}
\begin{flalign}
 \label{eqn:CAR1}\Psi_\mu(\overline{f}{}^\mu)^\dagger &= \bPsi^\mu(f_\mu)~,\\
 \label{eqn:CAR2}\big\lbrace \Psi_\mu(\overline{f}{}^\mu),\bPsi^\nu(h_\nu) \big\rbrace &= \langle f,h \rangle\,\mathsf{1}~,\\
 \label{eqn:CAR3}\big\lbrace \Psi_\mu(\overline{f}{}^\mu),\Psi_\nu(\overline{h}{}^\nu)\big\rbrace
&=\big\lbrace\bPsi^\mu(f_\mu),\bPsi^\nu(h_\nu)\big\rbrace =0~,
\end{flalign}
\end{subequations}
with $\dagger$ denoting the involution in the algebra.

As pointed out in \cite{Hack:2011yv},  non-negativity of the inner product is essential for the CAR (\ref{eqn:CAR}):
Assume any Hilbert space representation of the algebra above. Let $f_\mu\in\Sol$ be arbitrary 
and define $A:=\bPsi^\mu(f_\mu)$, then (\ref{eqn:CAR1}) and (\ref{eqn:CAR2}) imply
\begin{flalign}
 A^\dagger A +A A^\dagger = \langle f,f \rangle \,\mathsf{1}~.
\end{flalign}
From the expectation value in any normalized Hilbert space state $\vert \varphi\rangle$
one concludes
\begin{flalign}
\langle f,f\rangle =  \langle A \varphi\vert  A \varphi\rangle +\langle A^\dagger \varphi\vert  A^\dagger \varphi\rangle \geq 0~,
\end{flalign}
completing the argument.
As we have shown in Section~\ref{sec:innerproduct}, for $d$-dimensional FRW spacetimes
the inner product (\ref{eqn:innerproduct}) indeed satisfies
this necessary condition for a consistent quantization of the Rarita-Schwinger field.

The Dirac Rarita-Schwinger field as discussed above amounts to the generic case in $d$ dimensions
without imposing restrictions on $d$.
However, if a Majorana condition ${\bpsi^{}_\mu=\psi^\mathsf{T}_\mu C}$ is available and
used to reduce the Dirac spinor (e.g.\ in $d{=}4$ minimal supergravity),
the quantization proceeds in a similar way:
We restrict $\Sol$ to Majorana solutions $\Sol_\mathsf{maj}$ satisfying 
$\overline{f}{}_\mu^{} =f^\mathsf{T}_{\mu} C$.  
The inner product (\ref{eqn:innerproduct}) for Majorana solutions $f_\mu,h_\mu\in\Sol_\mathsf{maj}$ reads
\begin{flalign}
 \langle f,h \rangle
=-\int_\Sigma n^\mu f^{\nu\mathsf{T}}\, C\,\gamma_{\nu\mu\rho} \,h^\rho~.
\end{flalign}
It is symmetric since $C\gamma_{\nu\mu\rho}$ is anti-symmetric in the cases where a Majorana condition is available,
such that
\begin{flalign}
 h^{\nu\mathsf{T}} C\,\gamma_{\nu\mu\rho} \,f^\rho 
=f^{\rho\mathsf{T}} \big(C\gamma_{\nu\mu\rho}\big)^\mathsf{T} h^\nu
=f^{\rho\mathsf{T}} C\,\gamma_{\rho\mu\nu} h^\nu~,
\end{flalign}
and it is also real
\begin{flalign}
 \langle f,h \rangle^\ast = \langle h,f \rangle = \langle f,h\rangle~. 
\end{flalign}
We quantize the Majorana Rarita-Schwinger field in terms of a self-dual CAR-algebra \cite{Araki:1971id,Bar:2011iu}:
We associate to the Majorana solutions {\it hermitian} smeared field operators via the $\mathbb{R}$-linear map
$f_\mu\mapsto \bPsi_\mathrm{maj}^\mu(f_\mu)$. 
The self-dual CAR-algebra is defined as the $\ast$-algebra with unit $\mathsf{1}$ generated by these operators,
subject to the relations
\begin{flalign}
 \label{eqn:SCAR2}\big\lbrace \bPsi_\mathsf{maj}^\mu(f_\mu),\bPsi_\mathsf{maj}^\nu(h_\nu) \big\rbrace 
= \langle f,h \rangle\,\mathsf{1}~.
\end{flalign}

\section{\label{sec:causal}Causality and the role of supergravity}

In this section we discuss the propagation of the transversal and longitudinal parts 
of the Rarita-Schwinger field on $d$-dimensional FRW spacetimes \footnote{
We focus on the transversal and trace parts, $\psi_{m}^{\text{\,T}}$ and $\chi$, which are the
degrees of freedom entering the inner product (\ref{eqn:innerproductx}).
The explicit reconstruction of $\psi_m^{}$ via (\ref{eqn:decomposition2})
might be problematic due to the inverse powers of $k$.
We thank Thomas-Paul Hack for useful discussions on this issue.
}.
The relevant equations are the constraints (\ref{eqn:constraintsFRW}) 
and the equation of motion (\ref{eqn:gravitino-eom2}),
or equivalently (\ref{eqn:gravitino-eom2}), (\ref{eqn:constraint2FRW}) and (\ref{eqn:constraintderived}).
The non-dynamical $\psi_0^{}$ can be eliminated
by solving (\ref{eqn:constraint2FRW}), ${\psi^{}_0 = \gamma^{}_0\,\mathcal{A}\,\chi}$,
and (\ref{eqn:constraintderived}) is manifestly implemented in the decomposition (\ref{eqn:decomposition2}).
It thus remains to solve (\ref{eqn:gravitino-eom2}). 
The $\mu{=}0$ component yields the equation of motion for $\chi$
\begin{multline}\label{eqn:efflongprop}
  i \bigl(\gamma^0 \partial_0+ \gamma^{m}\mathcal{A}\,\partial_{m}\bigr)\chi
   +\frac{2\mathcal B -m}{d-2}\,\chi+\frac{d-1}{d-2}\,\mathcal B^{\,\dagger}\mathcal A\,\chi=0~.
\end{multline}
The spatial components of (\ref{eqn:gravitino-eom2}) give -- after using (\ref{eqn:efflongprop}) --
the equation for the transversal polarizations
\begin{flalign}\label{eqn:transprop}
i \gamma^\nu \partial_\nu \psi_m^\text{T} 
+ \left(\frac{i\,a^\prime}{2 a}(d-3)\gamma^0 -m\right) \psi_m^\text{T} =0~.
\end{flalign}
Thus, the transversal and longitudinal parts decouple.

Note that (\ref{eqn:transprop}) is a Dirac-type operator and therefore $\psi_m^\text{T}$ propagates causally, 
see e.g.~\cite{Bar:2011iu}.
In order to understand the causal properties of
the longitudinal part $\chi$, we define the `effective gamma matrices'
\begin{flalign}
 \gamma_\mathrm{eff}^0:=\gamma^0~,\quad \gamma_\mathrm{eff}^{m}:=\gamma^{m}\,\mathcal{A}~.
\end{flalign}
They form a Clifford algebra $\lbrace\gamma_\mathrm{eff}^\mu,\gamma_\mathrm{eff}^\nu \rbrace=2\,g_\mathrm{eff}^{\mu\nu}$
with an `effective metric' $g_\mathrm{eff}$ with components $g_\mathrm{eff}^{0\mu}=g^{0\mu}=a^{-2}\delta^{0\mu}$ and
\begin{flalign}\label{eqn:ceff}
g_\mathrm{eff}^{mn} = \bigl(\mathcal{A}_1^2 + a^{-2}\,\mathcal{A}_2^2\bigr)\, g^{mn}=: c_\mathrm{eff}^2(\tau)\, g^{mn}~.
\end{flalign}
The numerical coefficients $\mathcal{A}_{1}$ and $\mathcal{A}_{2}$ are defined by
 ${\mathcal{A}=:\mathcal{A}_1 + i\,\mathcal{A}_2\,\gamma^0}$, see (\ref{eqn:constraint2FRW}). 
Thus, (\ref{eqn:efflongprop}) is a Dirac-type operator on the spacetime with `effective metric' $g_\mathrm{eff}^{\mu\nu}$
and $\chi$ propagates causally with respect to $g_\mathrm{eff}^{\mu\nu}$.
Interpreted with respect to the original metric it
propagates with a time-dependent speed of light $c_\mathrm{eff}^2(\tau)$, as can be seen from (\ref{eqn:ceff}).
The propagation is therefore causal with respect to $g_{\mu\nu}$ as long as 
$c_\mathrm{eff}^2(\tau)\leq 1$ for all times $\tau$.
Note that any time-dependence of the mass leads to a positive contribution to the effective speed of light
since $\mathcal{A}_2^2 \propto (m^\prime)^2$.
Thus, time-variations in the mass can never reconcile an otherwise acausal propagation
of the longitudinal part with causality.

It is remarkable that the supergravity model discussed in \cite{Kallosh:1999jj,Kofman:1999ip,Kallosh:2000ve}
leads to $c_\mathrm{eff}^2(\tau)\equiv 1$ for all $d{=}4$ FRW solutions.
This ensures causal propagation of the gravitino on the one hand, but on the
other hand also means that 
the time-varying mass $m=e^{K/2} W$ exactly compensates the deficit in the effective
speed of light from being one, thus leading to standard causal properties.
As we will discuss in the next section, 
also for the Rarita-Schwinger field alone,
without the restrictions imposed by the supergravity model,
a causal propagation is possible 
on a variety of FRW spacetimes.
This, however, generically involves a time-dependent speed of light $c_\mathrm{eff}^2(\tau)\leq 1$.

\section{\label{sec:examples} Cosmological spacetimes}

Within the class of matter models described by the equation of state $p=\omega\rho$, $\omega\in \mathbb{R}$,
we identify the $d{=}4$ FRW spacetimes on which the Rarita-Schwinger field can propagate causally,
and study the time-dependence of the effective speed of light.
For a clearer physical interpretation we work in cosmological time $t$ defined by $dt=a(\tau) d\tau$,
such that the FRW metric reads $ds^2 = dt^2 - a(t)^2\,d\vec{x}^{\,2}$.
As discussed in the previous section, time-variations of the mass give a positive contribution
to the effective speed of light and thus
can only tighten the restrictions on the background spacetime.
We therefore focus on a {\it constant} mass $m>0$ and identify the spacetimes for which 
$c_\mathrm{eff}^2(t)\leq 1$ for all $t$.

\begin{figure}[htb]
 \includegraphics[width=0.9\linewidth]{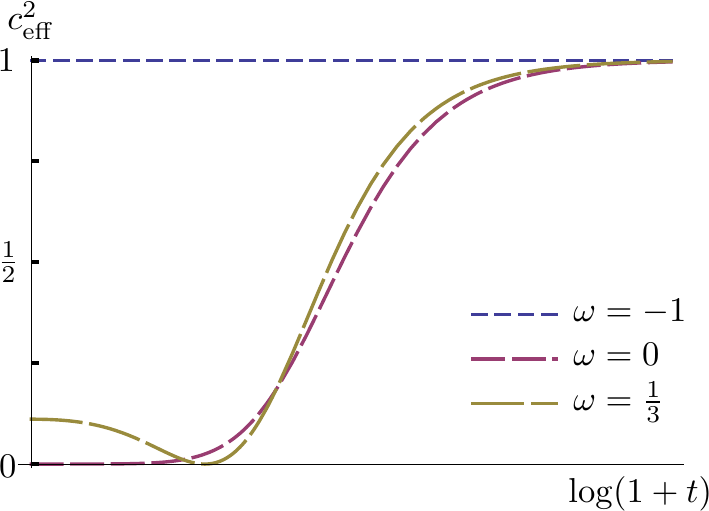}
\caption{\label{fig:plot} 
 Cosmological-time dependence of the effective speed of light for the longitudinal part.
 The plot shows with increasing dash length $\omega\in\lbrace -1,0,\frac{1}{3}\rbrace$,
 corresponding to a cosmological constant, dust and radiation dominated FRW universe, respectively.
}
\end{figure}

Solving Friedmann's equations for the Hubble rate $H$ yields
\begin{flalign}
 H(t):=\frac{\dot{a}(t)}{a(t)} = \frac{2}{3 t (\omega+1) + 2\alpha}~
\end{flalign}
with constant of integration $\alpha = H(0)^{-1}$.
The energy density is given by $\rho = 3 H^2$ and
the effective speed of light (\ref{eqn:ceff}) for the longitudinal part becomes
\begin{flalign}\label{eqn:ceff2}
 c_\mathrm{eff}^2(t) = \left(\frac{m^2 -\omega\,H(t)^2}{m^2 + H(t)^2 }\right)^2~.
\end{flalign}
For the special case $\omega=-1$, i.e.~de Sitter space,
we find $c_\mathrm{eff}^2(t)\equiv 1$ such that the Rarita-Schwinger field propagates 
with the standard speed of light, as expected.
Consider now the case $\omega\neq -1$, where we set $\alpha=0$ such that the cosmological singularity is
at $t=0$. For $t\to \pm\infty$ the Hubble rate vanishes and the speed of light $c_\mathrm{eff}^2$
approaches $1$. 
Thus, we find standard causal properties at late times.
On the other hand, for $t\to 0$ the Hubble rate diverges, such that $c_\mathrm{eff}^2\rightarrow\omega^2$.
For a causal propagation at times close to $t=0$ we have to require $\omega\in[-1,1]$.
In fact, from (\ref{eqn:ceff2})  this condition is necessary and sufficient for causal propagation
\begin{flalign}
 c_\mathrm{eff}^2(t)\leq 1  ~~\text{for all $t$}\quad \Longleftrightarrow\quad \omega\in[-1,1]~.
\end{flalign}
Interestingly, the matter models used in standard cosmology satisfy $\omega\in[-1,1]$
and hence allow for a causal propagation of the Rarita-Schwinger field.
We plot the effective speed of light for the cases ${\omega=-1}$ (cosmological constant),
${\omega=0}$ (dust) and ${\omega=1/3}$ (radiation) in Figure~\ref{fig:plot}.
Note that for $\omega>0$ the effective speed of light vanishes at  
$t=\pm 2\sqrt{\omega}/(3 m\,\vert \omega+1\vert)$.
This means that the longitudinal part of the Rarita-Schwinger field effectively does not propagate 
over extended spatial distances around these times.


\section{\label{sec:conc}Concluding remarks}
We have investigated the massive spin $3/2$ Rarita-Schwinger field 
focusing on the properties relevant for quantization.
Using the variational bicomplex we have constructed for 
generic spacetimes of dimension $d\geq 3$ a current which is conserved on solutions of the equation of motion. 
Furthermore, on $d$-dimensional FRW spacetimes we have shown that the associated inner product is 
positive definite and therefore allows for a consistent quantization of the Rarita-Schwinger
field in terms of a CAR-algebra.
This shows that also on non-Einstein spaces a consistent quantization is possible,
even without employing the additional structure of a supergravity.
We then have studied the propagation of the transversal and longitudinal parts of the 
Rarita-Schwinger field and found that, while the 
transversal polarizations propagate causally on all FRW spacetimes,
the propagation of the longitudinal part has quite distinct features.
Its propagation is characterized by a time-dependent effective speed of light,
and demanding causality imposes restrictions on the background spacetime and on time-variations of the mass.
This discussion offered an interesting perspective on the role of the time-dependent mass 
in the supergravity model \cite{Kallosh:1999jj,Kofman:1999ip,Kallosh:2000ve}.
For a constant mass we have found that the propagation is causal for $d{=}4$ FRW spacetimes
with a matter model described by the equation of state $p=\omega\,\rho$, if and only if $\omega\in[-1,1]$.
This in particular includes cosmological constant, dust and radiation dominated universes.
Comparing this result to the weak-field condition found for the electromagnetic background in \cite{Velo:1969bt}
which singles out preferred frames, 
we note that our condition is invariant under the FRW isometries.

The distinct features of the propagation of the longitudinal modes with 
time-dependent speed of light may also be relevant for models with explicit supersymmetry breaking, 
e.g.\ the  MSSM.
Interesting physical consequences may therefore be expected e.g.\ for bounds on gravitino dark matter.


\appendix

\section{\label{app:notation}Notation}
As usual, letters from the middle of the greek/beginning of the latin alphabet 
denote spacetime/flat Lorentz indices.
The metric signature is $(+,-,\dots,-)$ and $\eta_{ab}$ is the flat Minkowski metric.
The inverse vielbein is $e_a^\mu$, and $e:=\text{det}(e_\mu^a)$ is the volume element.
The Hodge operator is defined by $\star(dx^{\mu_1}\wedge \dots \wedge dx^{\mu_r}) = \frac{e}{(d-r)!} 
\epsilon^{\mu_1\dots\mu_r}_{\hphantom{\mu_1\dots\mu_r}\nu_{r+1}\dots \nu_d} dx^{\nu_{r+1}}\wedge\dots\wedge dx^{\nu_d}$.
The (flat) Dirac matrices satisfy the Clifford algebra $\lbrace\gamma^a,\gamma^b\rbrace =2\eta^{ab}$
and $\gamma^\mu:=e^\mu_a \gamma^a$.
Antisymmetrized products (of weight one) 
of gamma matrices are denoted by $\gamma^{\mu_1\dots \mu_n} := \gamma^{[\mu_1}\cdots\gamma^{\mu_n]}$.
The contraction of $\gamma^\mu$ with the Rarita-Schwinger field is denoted by
$\gamma\cdot \psi := \gamma^\mu\,\psi_\mu$.
The covariant derivatives of the vielbein and Rarita-Schwinger fields are
$\D_\mu e_\nu^a = \partial_\mu e_\nu^a +\omega_{\mu\hphantom{a}b}^{\hphantom{\mu}a}e_\mu^b- \Gamma_{\mu\nu}^\rho e_\rho^a $
and
$\D_\mu\psi_\nu =
 \partial_\mu\psi_\nu +\frac{1}{4}\omega_{\mu ab}\gamma^{ab} \psi_\mu-\Gamma_{\mu\nu}^\rho\psi_\rho$, 
$\D_\mu\bpsi_\mu=\overline{\D_\mu\psi_\mu}$.
The vielbein postulate $\D_\mu e_\nu^a=0$ relates the spin connection to the Christoffel symbols.
The covariant derivative of the gamma matrices reads 
$\D_\mu\gamma^\nu = \partial_\mu\gamma^\nu +\Gamma_{\mu\rho}^\nu\gamma^\rho + \frac{1}{4} \omega_{\mu ab}[\gamma^{ab},\gamma^\nu]$
and due to the vielbein postulate we have $\D_\mu\gamma^{\nu_1\dots \nu_n} =0$.


\begin{acknowledgments}
We thank Thomas-Paul Hack, Thorsten Ohl and also Julian Adamek and Florian Staub for useful discussions and comments.
CFU is supported by the German National Academic Foundation 
(Studienstiftung des deutschen Volkes).
AS and CFU are supported by Deutsche
Forschungsgemeinschaft through the Research Training Group GRK\,1147 
\textit{Theoretical Astrophysics and Particle Physics}.
\end{acknowledgments}


\bibliography{bibl.bib}

\end{document}